\documentclass[11pt]{tibop-article}

\TIBauthor[Valko et al.]{Danila Valko\affOne\affTwo\orcid{0000-0002-8058-7539}\\
    \small{\texttt{danila.valko@offis.de}}
\and\\{Jan S\"oren} Schwarz\affOne\affTwo\orcid{0000-0003-0261-4412}\\
    \small{\texttt{schwarz@offis.de}}
\and\\Jorge {Marx G\'omez}\affOne\orcid{0000-0002-7833-7549}\\
    \small{\texttt{jorge.marx.gomez@uol.de}}
\lastand\\Ralf Isenmann\affThree\orcid{0000-0002-1611-4892}\\
    \small{\texttt{ralf.isenmann@wb-fernstudium.de}}
}

\TIBaffiliations{\affOne Carl von Ossietzky Universit\"at Oldenburg, Ammerl\"ander Heerstraße 114-118, Oldenburg, Germany
\\\affTwo OFFIS – Institute for Information Technology, Escherweg 2, Oldenburg, Germany
\\\affThree Wilhelm Büchner Hochschule, Hilpertstraße 31, Darmstadt, Germany }

\TIBtitle{FAIR+S: A validation study of a framework for sustainable research data and software}
\TIBbundlename[PREPRINT]{}
\TIBabstract{The FAIR principles (Findable, Accessible, Interoperable, Reusable) have transformed research data management, but they do not address the environmental impact of creating and using research software and data, such as energy consumption, carbon emissions, and life‑cycle impacts that become central to computer science and engineering-related domains. To bridge this gap FAIR+Sustain\-ability or FAIR+S, an extension of the FAIR framework that embeds environmental accountability as a core element, was introduced.

Because FAIR principles already structure how digital research artefacts are described, shared, and reused, they offer an effective entry point for embedding sustainability considerations at scale. FAIR+S weaves carbon‑footprint and energy‑use considerations directly into FAIR‑aligned metadata schemas, workflows and development specifications. In doing so, it enables research infrastructures to report, compare, and audit the environmental implications of data and software in a measurable, interoperable, and transparent manner. This creates a foundation for reproducible research that simultaneously advances open science goals and decarbonisation objectives.

However, integrating environmental accountability into established research workflows raises questions of feasibility, relevance, and acceptance across stakeholders and disciplines. In this work we validated the framework through a cross-disciplinary expert survey. The evaluation confirms its importance and practical relevance, but also reveals current gaps in researchers' awareness of green software practices.
}
\TIBkeywords{\capitalisewords{FAIR, FAIR4RS, FAIR+S, Sustainability, Green Software Principles}}
\usepackage{xfrac}
\usepackage{mfirstuc-english}
\usepackage[export]{adjustbox}
\usepackage{longtable}
\usepackage{array}
\usepackage{tabularx}

\newcolumntype{L}[1]{>{\raggedright\arraybackslash}p{#1}}
\usepackage[flushleft]{threeparttable}

\usepackage{subcaption}
\usepackage{pbox}

\newcounter{q}
\newcommand{\q}[1]{\refstepcounter{q}\label{#1}}

\begin{document}
\maketitle

\section{Introduction}

The FAIR+S framework extends the FAIR guiding principles\cite{Wilkinson2016, Barker2022, Lamprecht2020} for research data and software management (RDSM) by explicitly integrating sustainability considerations into the lifecycle of digital research artefacts. While the FAIR principles have become widely adopted to improve data stewardship and reuse, they do not explicitly address the environmental impacts associated with research data and software lifecycle\cite{ResearchSoftwareLifecycle}, e.g. the production, execution, and long-term maintenance. FAIR+S addresses this gap by introducing sustainability-oriented principles that promote transparency, measurability, contextualised reproducibility, standardisation, and lifecycle responsibility for energy and resource use in digital research artefacts. The framework responds to growing evidence that computational research and digital research infrastructures contribute significantly to energy consumption and carbon emissions, and that sustainability-aware practices are increasingly necessary to support responsible software development\cite{Duboc2020, Becker2015} and reproducible research\cite{Lamprecht2020, DFG2025, Lannelongue2021}.

While these principles are conceptually compelling, their perceived importance by researchers and practical relevance cannot be directly assumed. As an extension intended for broad adoption, FAIR+S requires systematic validation to ensure that its principles are conceptually sound, practically feasible, and aligned with the expectations and constraints of its intended stakeholders. Without empirical validation, sustainability-oriented frameworks risk remaining aspirational or fragmented, limiting their effectiveness and uptake. The purpose of the present validation study is therefore to evaluate the relevance, feasibility, and perceived value of the FAIR+S principles through expert assessment, and to identify conditions under which they can be reliably implemented in real-world research and software development contexts. By grounding FAIR+S in expert agreement and empirical evidence, this study aims to strengthen its credibility as a sound and flexible framework that can inform community standards, research infrastructure practices, and policy development. 

We further posit that a rigorously validated FAIR+S framework is of interest not only to the academic research community but also to practitioners in industry and competent actors in agencies involved in data-intensive and software-driven innovation. As sustainability reporting, energy efficiency, and transparency become increasingly central to industrial research, digital services, and regulatory compliance\cite{Duboc2020, Becker2015, Hankendi2025, Demartini2025,}, FAIR+S offers a structured and interoperable approach for aligning scientific best practices with emerging industrial and policy requirements\cite{Molnar2025}. In this sense, FAIR+S has the potential to serve as a shared reference framework across research and industry, supporting sustainable digital practices while reinforcing the principles of openness, reproducibility, and long-term value creation.

\section{The FAIR+S Framework}
This section presents an updated and substantially extended description of the FAIR+S principles, which were first introduced at the 3rd NFDI4Energy Conference (24--25 March 2026) and later published in abbreviated form in the conference abstracts, see\cite{ValkoFAIRSNFDI4Energy}.

\subsection{Framework Aim and Structure}

To contextualise the subsequent validation, this section outlines the conceptual structure and scope of the FAIR+S framework\cite{ValkoFAIRSNFDI4Energy, ValkoFAIRS}. FAIR+S is designed as a complementary extension of the original FAIR\cite{Wilkinson2016} and FAIR4RS (FAIR for Research Software)\cite{Barker2022, Lamprecht2020} principles, preserving their core objectives while broadening their scope to address sustainability concerns\cite{ValkoFAIRSNFDI4Energy}. Rather than redefining Findability, Accessibility, Interoperability, or Reusability, FAIR+S builds on existing FAIR-aligned practices by introducing environmental accountability requirements that are orthogonal but interoperable with them (see Table~\ref{tab:fair_plus_s_longtable}). This alignment ensures conceptual continuity with established FAIR(4RS) initiatives, facilitates integration into existing data and software stewardship workflows, and avoids fragmenting standards ecosystems. By embedding environmental sustainability as an explicit dimension of responsible reuse, FAIR+S strengthens FAIR's long-term relevance in the context of environmentally accountable and policy-aligned open science.

\begin{longtable}{|L{3cm}|L{6.2cm}|L{6.2cm}|}
\caption{FAIR, FAIR4RS, and FAIR+S Guiding Principles}
\label{tab:fair_plus_s_longtable} \\
\hline
\textbf{Aspect} & \textbf{FAIR Principles}\cite{Wilkinson2016} & \textbf{FAIR4RS Principles}\cite{Barker2022} \\
\hline
\endfirsthead

\hline
\textbf{Aspect} & \multicolumn{2}{|L{12.4cm}|}{\textbf{FAIR+S Principles Extension}\cite{ValkoFAIRSNFDI4Energy}} \\
\hline
\endhead

\hline
\endfoot

\hline
\endlastfoot

\textbf{Findable} &
F1. (Meta)data are assigned a globally unique and persistent identifier. \newline
F2. Data are described with rich metadata. \newline
F3. Metadata clearly and explicitly include the identifier of the data they describe. \newline
F4. (Meta)data are registered or indexed in a searchable resource.
&
F1. Software is assigned a globally unique and persistent identifier. \newline
F1.1. Components representing different levels of granularity are assigned distinct identifiers. \newline
F1.2. Different versions of the software are assigned distinct identifiers. \newline
F2. Software is described with rich metadata. \newline
F3. Metadata clearly and explicitly include the identifier of the software they describe. \newline
F4. Metadata are FAIR, searchable, and indexable.
\\
\hline

\textbf{Accessible} &
A1. (Meta)data are retrievable by their identifier using a standardised communications protocol. \newline
A1.1. The protocol is open, free, and universally implementable. \newline
A1.2. The protocol allows for authentication and authorisation where necessary. \newline
A2. Metadata remain accessible even when the data are no longer available.
&
A1. Software is retrievable by its identifier using a standardised communications protocol. \newline
A1.1. The protocol is open, free, and universally implementable. \newline
A1.2. The protocol allows for authentication and authorisation where necessary. \newline
A2. Metadata remain accessible even when the software is no longer available.
\\
\hline

\textbf{Interoperable} &
I1. (Meta)data use a formal, accessible, shared, and broadly applicable language for knowledge representation. \newline
I2. (Meta)data use vocabularies that follow FAIR principles. \newline
I3. (Meta)data include qualified references to other (meta)data.
&
I1. Software reads, writes, and exchanges data in ways that meet domain-relevant community standards. \newline
I2. Software includes qualified references to other objects.
\\
\hline

\textbf{Reusable} &
R1. (Meta)data are richly described with accurate and relevant attributes. \newline
R1.1. (Meta)data are released with a clear and accessible license. \newline
R1.2. (Meta)data are associated with detailed provenance. \newline
R1.3. (Meta)data meet domain-relevant community standards.
&
R1. Software is described with a plurality of accurate and relevant attributes. \newline
R1.1. Software is given a clear and accessible license. \newline
R1.2. Software is associated with detailed provenance. \newline
R2. Software includes qualified references to other software. \newline
R3. Software meets domain-relevant community standards.
\\
\hline

\textbf{Sustainable} & \multicolumn{2}{|L{12.4cm}|}{
S1. Research artefacts should be described with accurate and relevant energy efficiency attributes, including measurements of resource use and carbon footprint. \newline
S2. Research artefacts and evaluation should be supplemented with community-accepted sustainability benchmarks. \newline
S3. Research artefacts, their development, and evaluation processes should align with existing sustainability frameworks and community standards where possible. \newline
S4. Carbon-related (meta)data and estimates must include transparency and accountability, disclosing uncertainty ranges, methodological assumptions, tools used, and responsible parties. \newline
S5. Research artefacts should, where possible, include information about their long-term life cycle sustainability, including expected maintenance, update frequency, and cumulative resource needs across development, deployment, and preservation.
}
\\
\hline
\end{longtable}

\subsection{Description of the Principles}

The following principles (S1–S5) operationalise the framework by specifying how sustainability considerations can be systematically integrated into RDSM and the research artefact lifecycle\cite{ResearchSoftwareLifecycle}. They bridge the green research software and data agenda with broader sustainability-oriented software initiatives, such as the Computing's Responsibility Manifesto\cite{Knowles2025}, the Karlskrona Manifesto\cite{Becker2015} and related sustainability awareness requirements\cite{Duboc2020, ValkoFAIRSNFDI4Energy}.

\subsubsection*{S1 – Energy Efficiency Attributes}

\textit{Research artefacts should be described with accurate and relevant energy efficiency attributes, including measurements of resource use and carbon footprint.}

Embedding such energy-related metadata within research artefacts allows both researchers and infrastructure providers to assess the environmental impact of their workflows and to make informed choices toward more energy-efficient alternatives.

Operationalising energy transparency requires tooling that is both sufficiently precise and compatible with existing research workflows. Despite existing energy evaluation tools, including Energy-Estimation Toolkits and PowerAPI\cite{bourdon, Fieni2024}, are somehow fragmented, they already enable precise enough measurement of energy performance across hardware platforms. Tools such as CodeCarbon\cite{codecarbon}, CarbonTracker\cite{anthony2021carbontracker_2}, Green Algorithms calculator\cite{Lannelongue2021}, and MLCO2\cite{MLCO2_18} streamline the process by automatically monitoring and estimating carbon emissions in computational workflows, while GreenAI\cite{schwartz2020green_25} initiatives advocating integration of energy consumption metrics directly into analytical and development pipelines to facilitate sustainable practices. 

\subsubsection*{S2 – Sustainability Benchmarks}

\textit{Research artefacts and evaluation should be supplemented with community-accepted sustainability benchmarks.} 

While energy metrics enable measurement at the artefact level, meaningful interpretation requires comparability across contexts and domains. Therefore sustainability reporting should rely on community-accepted benchmarks to ensure transparency and comparability. Such benchmarks provide standardised measurement methodologies, enabling sustainability assessments to be conducted consistently across projects, domains, and infrastructures. 

Community initiatives such as NFDI4Energy\cite{NFDI4Energy2023} are expected to play a key role in developing standardised sustainability benchmarks for evaluating the energy efficiency of computational infrastructures and data management processes, thereby fostering reproducible and comparable sustainability assessments. Similar to established corporate-level\cite{Rekker2025} or domain-oriented benchmarks, such as MLPerf\cite{reddi2019mlperf} for machine learning workloads and SPEC\-power\cite{SPEC} for system-level energy efficiency, these community-driven efforts, aligned with the FAIR+S principles, could support the creation of shared reference models and configurations to promote environmental accountability at systemic level.

\subsubsection*{S3 – Alignment with Sustainability Frameworks}

\textit{Research artefacts, their development, and evaluation processes should, where possible, align with existing sustainability frameworks and community standards.} 

Beyond benchmarks for comparison, FAIR+S emphasises alignment with broader sustainability frameworks that govern environmental accounting and reporting. Aligning research and software development practices with recognised sustainability frameworks and standards, such as ISO/IEC 21031:2024\cite{GSF2023}, W3C Sustainability Guidelines\cite{w3c}, Sustainability awareness framework\cite{Duboc2020}, the Greenhouse Gas Protocol\cite{Ranganathan, ghg}, and ISO\-14044\cite{ISO14044} for life cycle assessment, enhances credibility, transparency, and interoperability between academic and industrial software and data ecosystems. These frameworks and standards are supposed to provide complementary guidance for evaluating environmental impacts, reporting emissions, and ensuring methodological consistency across sustainability assessments in different contexts. Adherence to such foundational frameworks is a core principle of sound environmental accounting within the FAIR+S para\-digm. One notable community initiative is the Green DiSC certification\cite{sysinst} for research labs and teams -- an effort led by the Software Sustainability Institute to provide computational researchers and research-performing organisations with a clear roadmap for addressing and reducing the environmental impacts of their work.

\subsubsection*{S4 – Carbon Transparency and Accountability}

\textit{Carbon-related metadata and estimates must include transparency and accountability, disclosing uncertainty ranges, methodological assumptions, tools used, and responsible parties}.

Alignment alone is insufficient without mechanisms that ensure transparency and accountability in how sustainability claims are produced\cite{Becker2015}. This principle mandates the disclosure of critical metadata, such as hardware specifications, energy source mixes, and measurement methodologies, in accordance with open science best practices. In practice, this principle can be operationalised through digitalised frameworks such as digital Measuring, Reporting, and Verification (dMRV)\cite{Korner2025dMRV}.

Transparent reporting of measurement assumptions, hardware context, and associated uncertainties is essential to ensure that sustainability data remain reproducible, verifiable, and trustworthy. 

\subsubsection*{S5 – Life Cycle Sustainability}

\textit{Research artefacts should, where possible, include information about their long-term life cycle sustainability, including expected maintenance, update frequency, and cumulative resource needs across development, deployment, and preservation.} 

In addition to methodological transparency at a given point in time, sustainability must be evaluated across the full life cycle of digital research artefacts. FAIR+S expands sustainability assessment beyond runtime metrics to include maintenance, storage, and reuse over the full research artefact life cycle. This perspective supports responsible and reproducible science\cite{DFG2025, Lannelongue2021, Lamprecht2020} and aligns with ISO 14044 and Life Cycle Assessment (LCA) methodologies\cite{Falk2025, Nair2025}, encouraging sustainable long-term curation of datasets and software.

\section{Methods}

\subsection{Expert Survey Design and Recruitment Strategy}

To empirically assess the relevance, feasibility, and perceived value of FAIR+S, the framework was validated through a structured expert survey grounded in a Design Science Research\cite{hevner2004design,Peffers2007} evaluation approach (see full methodological details in \hyperref[sec:appendixB]{\textit{Appendix B}}). The methodology followed a systematic process including survey design based on expert elicitation principles, pilot testing, purposive expert recruitment with defined eligibility criteria, and a mixed-methods analytical strategy combining quantitative and qualitative analysis. 

The survey instrument was developed in accordance with established best practices in expert elicitation and survey design\cite{Rowe2001, Burgman2011, Bojke2021}. The main questions addressed (1) experts' awareness of and familiarity with state-of-the-art concepts related to the FAIR principles\cite{Wilkinson2016, Barker2022}, the Sustainable Development Goals (SDGs)\cite{SDGs}, and green software practices (GSP)\cite{GSF2023}, and (2) a structured introduction to FAIR+S, followed by an assessment of its relevance, feasibility, and applicability (see \hyperref[sec:appendixB]{\textit{Appendix B}}). A combination of Likert-scale items, ranking questions, and open-ended prompts was used to balance quantitative assessment with qualitative insight, while minimising respondent burden and supporting content validity\cite{Dillman2014}.

Participants were recruited between 2th February and 15th April 2026 using purposive sampling that is consistent with expert consensus methodologies\cite{Rowe2001, Burgman2011, Bojke2021} to ensure domain expertise, targeting researchers, data stewards, and software developers with demonstrable experience in RDSM, software development, sustainability assessment, or research governance. Recruitment was conducted through professional networks, academic mailing lists, and research infrastructure communities to promote disciplinary diversity. For instance, the survey invitation was disseminated via internal communication channels across OFFIS – Institute for Information Technology, the L3S Research Center, and the Carl von Ossietzky University of Oldenburg. In addition, selected expert communities were contacted directly (e.g., the Green Software Foundation (GSF), the National Research Data Infrastructure (NFDI), Software Sustainability Institute (SSI), Society of Research Software Engineering), while others were reached through sharing the invitation on LinkedIn. The survey was also distributed during the 3rd NFDI4Energy conference in 24-25 March 2026, which brought together experts in the fields of energy and data management\cite{ValkoFAIRSNFDI4Energy, NFDI4Energy2023}.

Participation in the survey was voluntary, informed consent was obtained digitally, and responses were collected anonymously in accordance with ethical guidelines. Ethical approval was obtained from the Carl von Ossietzky Universit\"at Oldenburg. The survey was piloted prior to deployment to refine question clarity and reliability.

\subsection{Expert Profile and Eligibility Criteria}

Given normative-driven nature of FAIR+S, careful expert selection was essential to ensure meaningful validation.

After the data collection deadline, a total of 40 valid responses were received. Eligibility criteria were then applied to select an expert sample for the validation part of this work (see, \ref{validation}), ensuring participants have relevant domain expertise and qualified experience\cite{Burgman2011, Bojke2021}.

Experts were selected if they met at least one of the following two primary criteria: (1) demonstrated familiarity with the FAIR principles, SDGs, or GSP; (2) a minimum of five years of research experience or having a PhD. In addition, participants were required to meet at least one of the following two experience-based criteria: (3) extensive experience in data collection or dataset preparation; (4) extensive experience in research software development or the use of programming languages for research.

This step resulted in a final sample of 27 experts, which is considered a sufficient number given that not all experts equally cover all domains of expertise\cite{Bojke2021, Budescu2014, Burgman2011} (see sample size remarks in \hyperref[sec:appendixB]{\textit{Appendix B}}). Although not all participants met the eligibility criteria, the full sample was retained to illustrate the adoption of FAIR principles among researchers and their readiness for FAIR+S in general (see, \ref{adoption}).

\begin{figure*}[t!]
    \begin{subfigure}[t]{0.45\textwidth}
        \includegraphics[scale=0.7, center]{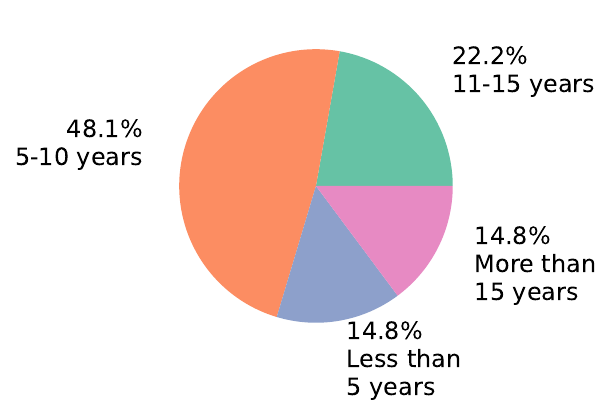}%
        \caption{Years of experience in research.}
    \end{subfigure}%
    \hspace{0.5cm} 
    \begin{subfigure}[t]{0.45\textwidth}
        \includegraphics[scale=0.7, center]{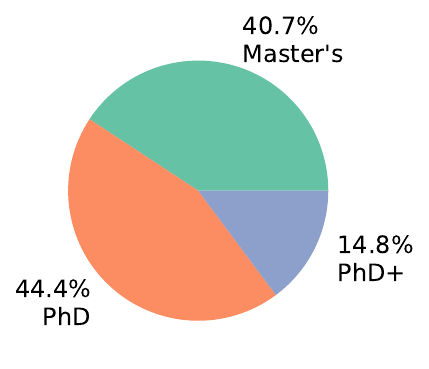}
        \caption{The level of education.}
    \end{subfigure}\vspace{1cm}
    ~
    \begin{subfigure}[t]{0.45\textwidth}
        \includegraphics[scale=0.7, center]{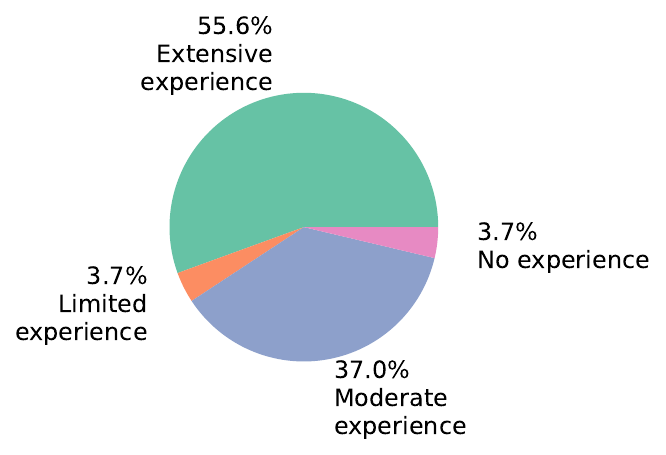}%
        \caption{Experience in data collection or preparation.}
    \end{subfigure}%
    \hspace{0.5cm} 
    \begin{subfigure}[t]{0.45\textwidth}
        \includegraphics[scale=0.7, center]{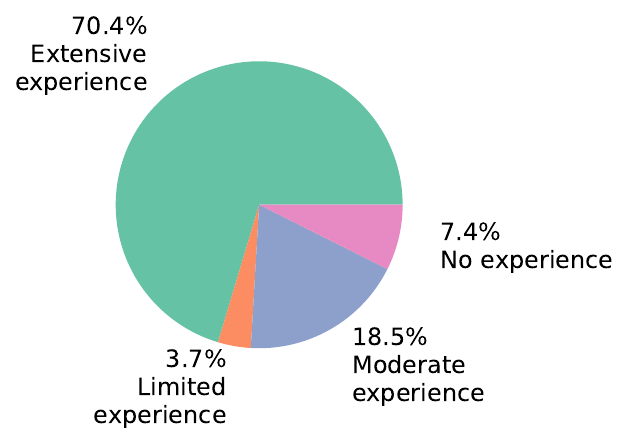}
        \caption{Experience in developing software.}
    \end{subfigure}

    \caption{The expert panel characteristics.}
    \label{fig01}

\begin{tablenotes}[]\footnotesize\item[]Note. $n = 27$, representing the number of complete respondent replies used in this figure.
\end{tablenotes}
    
\end{figure*}

The Fig.~\ref{fig01} summarises the background (see, Q\ref{q02}--Q\ref{q06} in \hyperref[sec:appendixA]{\textit{Appendix A}}) of the expert panel used in the validation part of study. In terms of research experience (Fig.~\ref{fig01}a), most experts have more than five years of experience, with 48.1\% reporting 5–10 years, 22.2\% reporting 11–15 years, and 14.8\% having more than 15 years; only 14.8\% have less than five years of experience. Regarding formal education (Fig.~\ref{fig01}b), all selected panel members hold advanced degrees: 40.7\% have a Master's degree, 44.4\% hold a PhD, and 14.8\% have PhD+ qualifications (e.g., professor or group leader).

Experience in data-related work is also strong (Fig.~\ref{fig01}c), with 55.6\% reporting extensive experience in data collection or dataset preparation and 37.0\% moderate experience. Similarly, 70.4\% of participants report extensive experience in software development (Fig.~\ref{fig01}d), with 18.5\% reporting moderate experience. Overall, these results indicate that the panel is well qualified, with either substantial research or technical expertise.

\begin{figure*}[!ht]
\centering
\includegraphics[scale=0.8, center]{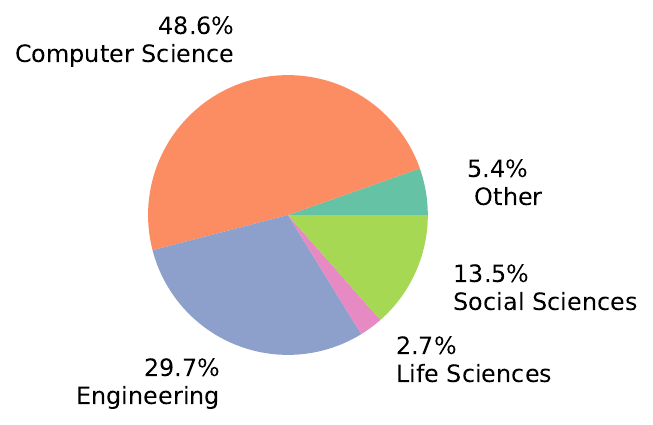}
\caption{The expert's main area of research.}\label{fig02}

\begin{tablenotes}[]\footnotesize\item[]Note. $n = 27$, representing the number of complete respondent replies used in this figure.
\end{tablenotes}

\end{figure*}

Experts mainly chose computer science and engineering as their primary field of research, but social sciences and other fields are also presented (see Fig.~\ref{fig02} and Q\ref{q01} in \hyperref[sec:appendixA]{\textit{Appendix A}}).

In addition to practical experience, experts should have sufficient familiarity with relevant concepts, demonstrating their relevance to specific domains.

\begin{figure*}[!ht]
\centering
\includegraphics[scale=0.65, center]{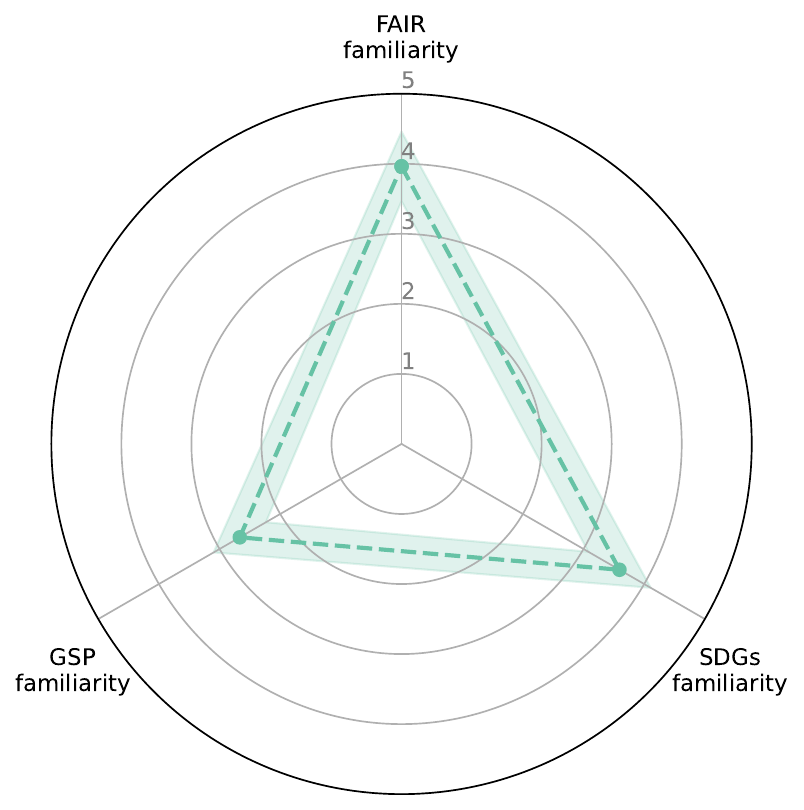}
\caption{Experts' familiarity with relevant concepts.}\label{fig03}

\begin{tablenotes}[]\footnotesize\item[]Note. Familiarity is measured on a five-point scale (1 = not familiar, 5 = very familiar). Average values are indicated by a dashed line, and the shaded area represents the 95\% confidence interval. $n = 27$, representing the number of complete respondent replies used in this figure.
\end{tablenotes}

\end{figure*}

Fig.~\ref{fig03} illustrates experts' familiarity with relevant concepts, measured on a five-point scale (1 = not familiar, 5 = very familiar) and reported as average values with the 95\%-confidence interval. The figure compares familiarity with FAIR (see, Q\ref{q07} in \hyperref[sec:appendixA]{\textit{Appendix A}}) to familiarity with broader sustainability-related frameworks such as the SDGs (see, Q\ref{q10}) and GSP (see, Q\ref{q11}). As shown, familiarity with FAIR ($M = 4.0$) and the SDGs ($M = 3.6$) is relatively higher than with GSP ($M = 2.7$), which is expected given that GSP are not yet widely adopted in research software and data development. This observation further indicates that the proposed framework is timely.

Overall, the Fig.~\ref{fig03} shows that experts report varying but generally moderate to high levels of familiarity across these concepts, demonstrating that their expertise extends beyond practical experience and is grounded in the conceptual knowledge required to validate the proposed framework.

\begin{figure*}[!ht]
\centering
\includegraphics[scale=0.75, center]{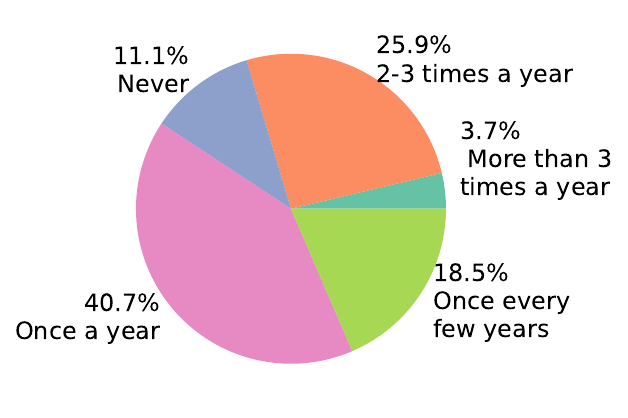}
\caption{The expert's review activities.}\label{fig04}

\begin{tablenotes}[]\footnotesize\item[]Note. $n = 27$, representing the number of complete respondent replies used in this figure.
\end{tablenotes}

\end{figure*}

Additionally, Fig.~\ref{fig04} shows that approximately 70.3\% of experts participate in paper or grant review activities at least once per year (see, Q\ref{q04}), indicating that they are likely aware of current open science standards and grant requirements.

Together, this ensured a panel capable of evaluating both conceptual soundness and practical feasibility, forming the basis for the analytical approach described next.

\subsection{Analytical Approach}

The evaluation of the FAIR+S framework was conducted using a mixed analytical approach combining descriptive, comparative, and limited content analyses to assess principle-level relevance, feasibility, and adoption readiness. 

Quantitative survey responses were analysed using descriptive statistics and distributional analyses to identify central tendencies and variability across expert judgments, following established practices for evaluating survey data\cite{Boone2012}. Relationships between perceived importance, feasibility, and adoption were examined through comparative analyses to identify alignment or tension\cite{Rowe2001}. 

Open-ended responses were analysed using qualitative content analysis to systematically identify recurring barriers, enablers, and refinement needs, supporting interpretive validation of the quantitative findings.

Together, these analytical methods enabled a elaborated evaluation of FAIR+S that integrates statistical evidence with expert interpretation to support robust conclusions about framework validity and adoption potential\cite{Yan2012, Steenbergen2007}.

\section{Main Results}

\subsection{Adoption of FAIR Principles and Readiness for FAIR+S}
\label{adoption}

\begin{figure*}[!ht]
        \includegraphics[scale=0.65]{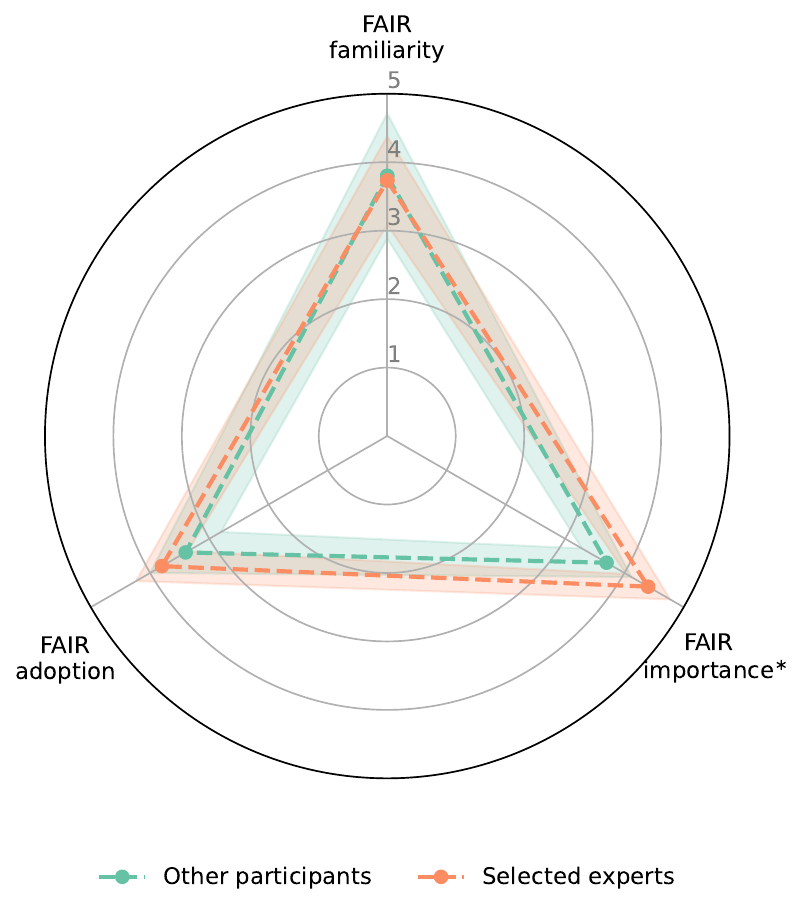}
    ~\\
        \small\begin{tabular}{p{8cm}cc}
        \hline
         & Other participants & Selected experts \\
        \hline
        How familiar are you with the FAIR principles? & 3.77 & 3.96 \\
        How important is it to follow FAIR principles? & 3.82 & 4.40 \\
        To what extent do you apply the FAIR principles? & 3.42 & 3.81 \\
        \hline
        \end{tabular}
    \caption{Importance and adoption of FAIR principles.}
    \label{fig05}

\begin{tablenotes}[]\footnotesize\item[]Note. Categories are measured on a five-point scale (1 = not familiar/important, never apply, 5 = very familiar/important, always apply). Average values are indicated by a dashed line, and the shaded area represents the 95\% confidence interval. $n_{\text{selected}} = 27$, $n_{\text{other}} = 13$, representing the number of complete respondent replies used in this figure. Significance levels for the Mann–Whitney U test are denoted as follows: *** $p < 0.001$, ** $p < 0.01$, * $p < 0.05$.
\end{tablenotes}
    
\end{figure*}

Before assessing FAIR+S directly, it is necessary to establish baseline adoption and acceptance of the underlying FAIR principles. Fig.~\ref{fig05} illustrates the reported importance (see, Q\ref{q08} in \hyperref[sec:appendixA]{\textit{Appendix A}}) and adoption (see, Q\ref{q09}) of the FAIR principles across the surveyed experts and provides an important baseline for assessing readiness for the FAIR+S framework.

Overall, all respondents, including selected experts and other researchers, engineers, and software developers, report high levels of familiarity with FAIR ($M_{selected} = 4.0$, $M_{other} = 3.8$, see, Fig.~\ref{fig05}) and consistently rate its importance as high or very high ($M_{selected} = 4.4$, $M_{other} = 3.8$), indicating broad conceptual acceptance of FAIR as a foundation for responsible RDSM. 

Reported by levels of FAIR adoption are slightly lower than perceived importance, reflecting a well-documented implementation gap between normative support and practical application\cite{Mons2017, FAIRsFAIR}. This pattern is observed across the full sample and is further accentuated when comparing with the selected expert subgroup ($M_{selected} = 3.8$, $M_{other} = 3.4$, see, Fig.~\ref{fig05}). 

At the same time, the overall statistical similarity in responses between experts and the other respondents suggests that FAIR principles are not restricted to a narrow specialist community but are widely recognised by professional roles. 

Notably, selected experts exhibit a statistically significant higher importance rate than the rest of respondents ($M = 4.4$ vs. $3.8$; Mann–Whitney $U$ test $p < 0.05$). This convergence, together with respondents’ reported knowledge of the SDGs and GSP (see, Fig.~\ref{fig03}), supports the view that FAIR+S can be introduced across the broader cohort as an incremental extension of existing FAIR(4RS) practices rather than as a fundamentally new paradigm. Nevertheless, the observed adoption gap highlights the need for additional guidance, tooling etc.

Thus this baseline provides the contextual foundation for evaluating how FAIR+S is perceived as an extension of existing practices.

\subsection{Framework Validation Results}
\label{validation}

The evaluation of FAIR+S focused on three complementary validation dimensions: perceived importance, added value, and practical feasibility. 

\subsubsection{Perceived Importance}

Perceived importance was assessed using principle-level importance ratings of the FAIR+S dimensions ($S1$–$S5$, see, Q\ref{q30} in \hyperref[sec:appendixA]{\textit{Appendix A}}).

\begin{figure*}[!ht]
\centering
\includegraphics[scale=0.65, center]{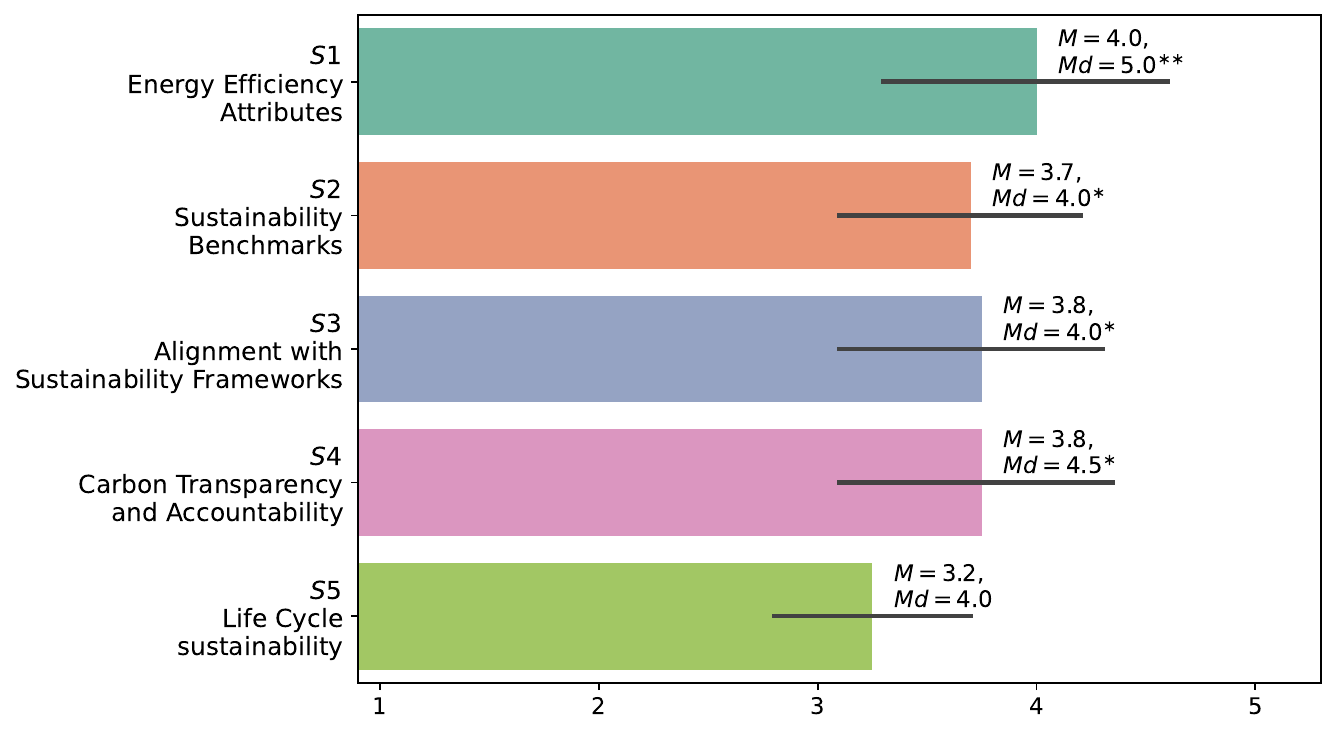}
\caption{FAIR+S principles importance.}\label{fig06}

\begin{tablenotes}[]\footnotesize\item[] Note. Importance is measured on a five-point scale (1 = not important, 5 = very important). Mean values are denoted by $M$, and medians by $Md$. 95\% confidence intervals for mean values are shown with error bars. $n = 20$, representing the number of complete respondent replies used in this figure. Significance levels are reported for the Wilcoxon signed-rank test against a median of 3: *** $p < 0.001$, ** $p < 0.01$, * $p < 0.05$.
\end{tablenotes}

\end{figure*}

Fig.~\ref{fig06} summarises the perceived importance of the FAIR+S principles across experts.

All principles are consistently rated from moderately important to very important on average ($M_{S1} = 4.0$, $M_{S2} = 3.7$, $M_{S3} = 3.8$, $M_{S4} = 3.8$, $M_{S5} = 3.2$). All mean ranks are significantly higher than 3, as indicated by the Wilcoxon signed-rank test (statistically significant for all principles except $S5$), suggesting broad endorsement of the FAIR+S framework.

Thus, experts consistently assigned high importance ratings across all principles, with little variation between items. However, due to the lack of differentiation in ratings, no conclusions can be drawn regarding the relative prioritisation of principles, as the Friedman test did not indicate significant differences between principle ranks ($\chi^2 = 8.5$, $p = 0.075$).

\subsubsection{Added Value}

Beyond perceived importance, validation also requires assessing whether FAIR+S is seen as delivering concrete benefits to research practice.

The added value of the framework was assessed through survey questions targeting perceived trust enhancement, transparency, and the usefulness of sustainability-related disclosures. Trust-related value was primarily evaluated by asking experts to assess the usefulness of standardised sustainability certifications for software and datasets (see, Q\ref{q22} in \hyperref[sec:appendixA]{\textit{Appendix A}}) and whether the inclusion of lifecycle sustainability statements would improve their trust in published research artefacts (see, Q\ref{q28}).

\begin{figure*}[!ht]

    \begin{subfigure}[t]{0.5\textwidth}
        \includegraphics[scale=0.65, center]{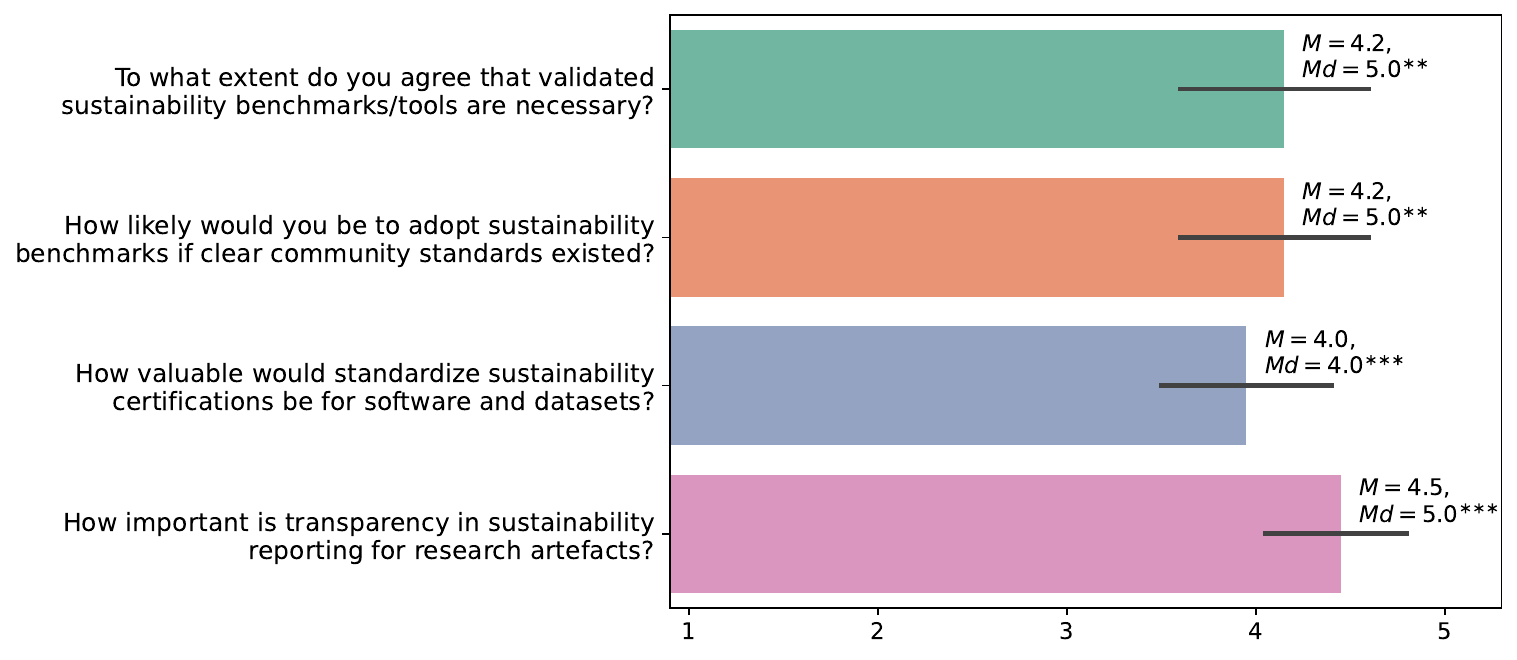}%
        \caption{Different aspects of added value.}
    \end{subfigure}%
    ~ \\
    \begin{subfigure}[t]{0.5\textwidth}
        \includegraphics[scale=0.5, center]{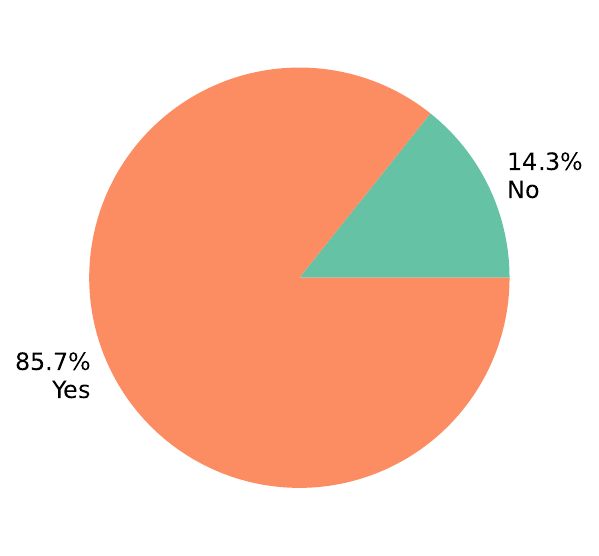}%
        \caption{Would lifecycle sustainability statements improve your trust in
published datasets/software?}
    \end{subfigure}%
    ~ 
    \begin{subfigure}[t]{0.5\textwidth}
        \includegraphics[scale=0.5, center]{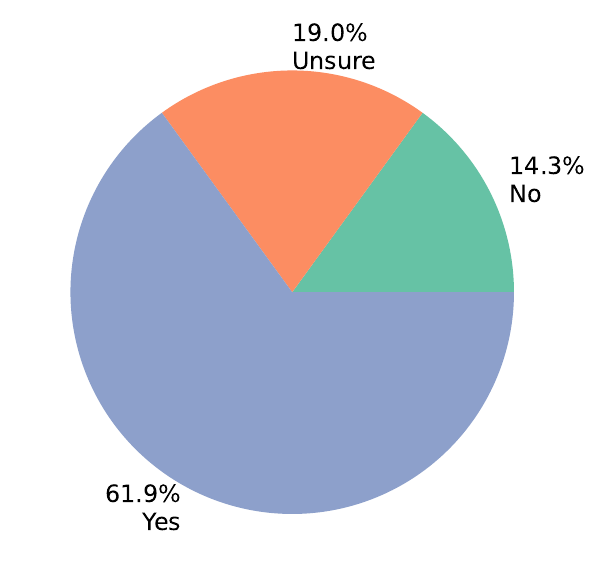}%
        \caption{Would you support
mandatory disclosure of
methods/tools used for energy
measurement?}
    \end{subfigure}%

\caption{FAIR+S principles added value.}\label{fig07}

\begin{tablenotes}[]\footnotesize\item[]Note. (a) Added value is measured on a five-point scale (1 = not important/likely/valuable, 5 = very important/likely/valuable). Mean values are denoted by $M$, and medians by $Md$. 95\% confidence intervals for mean values are shown with error bars, $n = 20$. Significance levels are reported for the Wilcoxon signed-rank test against a median of 3: *** $p < 0.001$, ** $p < 0.01$, * $p < 0.05$. (b, c) $n = 21$, representing the number of complete respondent replies used in this figure.
\end{tablenotes}

\end{figure*}

Experts rated standardised sustainability certifications for software and datasets as highly valuable, with responses strongly concentrated at the upper end of the scale (see, Fig.~\ref{fig07}a), indicating broad support for certification mechanisms analogous to FAIR badges.

This perception of enhanced trust was further reinforced by responses to Q\ref{q28}, where a majority of respondents (85.7\%, see, Fig.~\ref{fig07}b) indicated that lifecycle sustainability statements would improve their trust in published datasets and software. This result suggests that such statements are perceived as meaningful trust signals rather than superficial additions.

Transparency and accountability emerged as central dimensions of the framework's perceived added value. Transparency in sustainability reporting (see, Q\ref{q24}) was rated as extremely important by most experts, with responses again strongly skewed toward the highest levels of agreement (see, Fig.~\ref{fig07}a).

In parallel, support for mandatory disclosure of the methods and tools used for energy measurement and carbon estimation (see, Q\ref{q25}) was substantial. A clear majority of respondents (61.9\%, see, Fig.~\ref{fig07}c) expressed support, while only a small proportion opposed the requirement, indicating general acceptance of enforceable transparency measures despite some residual uncertainty.

Experts also largely agreed on the necessity of validated sustainability benchmarks and tools, reporting a high likelihood of adoption should clear community standards be established (see, Q\ref{q19}, Q\ref{q20}). 

Taken together, these findings suggest that FAIR+S is not perceived merely as an additional reporting burden. Rather, it is viewed as a framework capable of improving comparability, interpretability, and trust in data and software artefacts by operationalising environmental accountability in a standardised and transparent manner.

\subsubsection{Feasibility, Challenges and Concerns}

Despite this strong perceived added value, experts identified significant challenges to practical implementation.

Practical feasibility was examined through questions on the practicality of including sustainability metadata and reporting execution context (Q\ref{q13}, Q\ref{q14} in \hyperref[sec:appendixA]{\textit{Appendix A}}), complemented by self-reported current practices and anticipated barriers (Q\ref{q15}, Q\ref{q16}, Q\ref{q23}).

\begin{figure*}[!ht]

\begin{subfigure}[t]{0.5\textwidth}
        \includegraphics[scale=0.65, center]{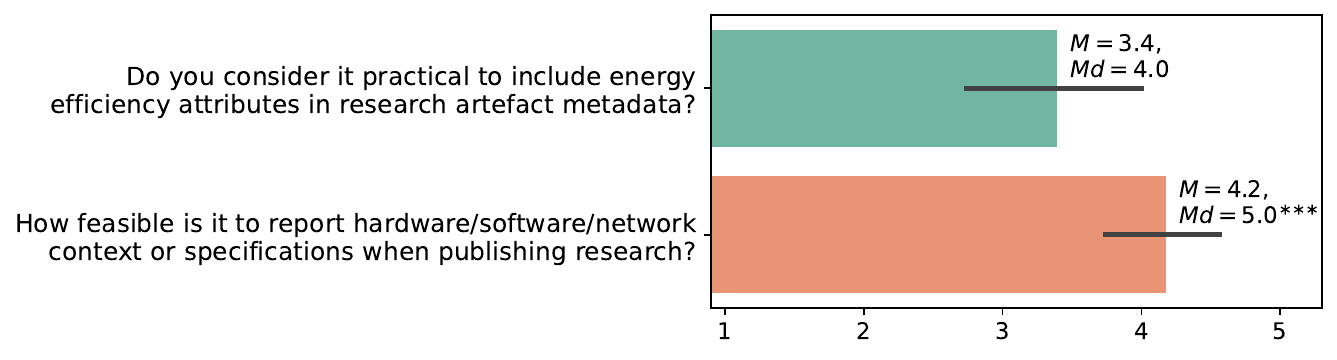}%
        \caption{Different aspects of feasibility.}
    \end{subfigure}%
    ~ \\

    \begin{subfigure}[t]{0.5\textwidth}
        \includegraphics[scale=0.65, center]{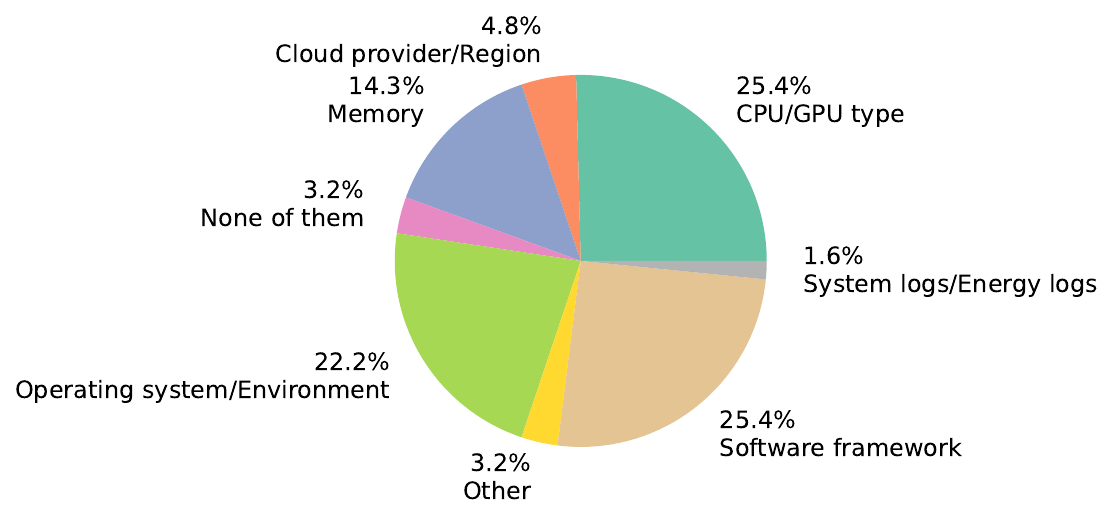}%
        \caption{Which hardware/software details do you currently report (if any)?}
    \end{subfigure}%

\caption{FAIR+S principles practical feasibility.}\label{fig08}

\begin{tablenotes}[]\footnotesize\item[]Note. (a) Feasibility is measured on a five-point scale (1 = not feasible/practical, 5 = very feasible/practical). Mean values are denoted by $M$, and medians by $Md$. 95\% confidence intervals for mean values are shown with error bars, $n = 23$. Significance levels are reported for the Wilcoxon signed-rank test against a median of 3: *** $p < 0.001$, ** $p < 0.01$, * $p < 0.05$. (b) $n = 27$, representing the number of complete respondent replies used in this figure.
\end{tablenotes}

\end{figure*}

While experts generally viewed sustainability reporting as feasible in principle (Fig.~\ref{fig08}a), feasibility ratings were consistently higher for execution context attributes than for energy efficiency attributes, and overall slightly lower than the perceived importance of transparency in this regard (see Fig.~\ref{fig07}a). This discrepancy suggests a clear implementation gap, likely driven less by conceptual resistance and more by practical limitations -- particularly the absence of standardised metrics, tools, and concrete guidance.

Notably, respondents already report many execution context attributes (Fig.~\ref{fig08}b), such as CPU/GPU type, memory, software frameworks, and operating system details, which are also essential for carbon footprint transparency.

From the perspective of anticipated challenges and barriers, expert feedback highlights several factors that may hinder effective adoption of the framework. A recurring concern is the additional time and effort required to report energy efficiency metrics. Researchers already operate under tight constraints, and the need to collect, estimate, and document supplementary information is widely perceived as an added burden, especially when such efforts are not explicitly rewarded or mandated by publication venues, funding bodies, or institutions.

Another prominent barrier relates to the lack of accessible and standardised tools. Many respondents emphasised that without automated, easy-to-use solutions that integrate seamlessly into existing research workflows, practical implementation remains difficult. Manual measurement and reporting are viewed as both error-prone and overly demanding, particularly given the diversity of hardware configurations, operating systems, and programming environments across research settings.

Limited expertise further compounds feasibility concerns. Researchers frequently report insufficient technical knowledge to measure energy consumption reliably or to translate raw measurements into comparable indicators such as carbon emissions. In the absence of clear methodological guidance, training materials, or built-in support, this knowledge gap undermines consistency and reduces confidence in the reported values.

Practical constraints of real-world research infrastructures also play a significant role. In shared environments such as high-performance computing clusters or cloud platforms, researchers often lack visibility into underlying hardware characteristics or energy usage, rendering direct measurement infeasible. 

Consequently, these challenges point not to the infeasibility of sustainability measurement and environmental accountability, but to the need for institution-level solutions in labs and university infrastructures. As illustrated by Microsoft's Azure Architecture guidance on measuring application sustainability\cite{msazure2026}, meaningful and comparable carbon assessment can be achieved through standardised metrics, estimation methods, and integration into existing workflows rather than through exhaustive, manual measurement at the individual researcher level. In parallel, a mature ecosystem of tools already exists that can operationalise this approach, including energy-estimation toolkits such as CodeCarbon\cite{codecarbon}, CarbonTracker\cite{anthony2021carbontracker_2}, and the Green Algorithms calculator\cite{Lannelongue2021}, which automate the monitoring or estimation of energy use and carbon emissions across diverse computational workflows. If adopted centrally and embedded into shared infrastructures such as HPC clusters, cloud environments, or institutional computation systems, these tools can directly address the lack of accessibility, reduce the burden of manual reporting, and mitigate expertise gaps by providing built-in methodological guidance. Universities and research labs are therefore well positioned to lower practical barriers by standardising tool support, offering training, and providing default sustainability metrics, enabling consistent and credible reporting even in heterogeneous, shared, or partially opaque research environments. This can be done by adapting sustainable organisation-level frameworks to university and research institution environments, like the Sustainable Organisational Framework for Technology (SOFT)\cite{soft2026}.

Of course, uncertainty regarding what should be reported and where system boundaries should be drawn further limits adoption. Ambiguity around relevant attributes, levels of granularity, and inclusion or exclusion of indirect factors may reduce potential applicability in each concrete case. At the same time, the lack of strong incentives, shared norms, or formal requirements makes it difficult for researchers to justify the added reporting burden, particularly when the perceived relevance to their specific research output is low. To overcome these challenges, a clear and practical path forward is the development of lightweight, standardised reporting practices supported by automated tools and minimal reporting requirements. This includes defining a small set of core metrics, providing default estimation methods where direct measurement is not possible, and embedding energy reporting into commonly used research workflows and infrastructures. Clear guidance, reference implementations, and alignment with publication and funding requirements would substantially lower adoption barriers while preserving scientific flexibility.

A significant step in this direction could be taken by the GSF (\url{https://greensoftware.foundation/}), the SSI (\url{https://www.software.ac.uk/}), and the NFDI4Energy initia\-tive\cite{NFDI4Energy2023}. The last' mission is already align well with these challenges and objectives\cite{NFDI4Energy_ProgressReport}. 
This could include integrating FAIR+S-aligned descriptors into the metadata schemas used and developed by NFDI4Energy repositories, enabling richer and more consistent sustainability metadata\cite{schmurr_2025_17549282,Hulk_Open_Energy_Family_2025,qussous_2026_18390186,ferenz2026metadataschemaenergyresearch}. 
Additionally, the Simulation-as-a-Service hub of NFDI4Energy can be used as use case for the exemplary implementation of FAIR+S in research infrastructures\cite{seiwerth_2025_15065996}.

Overall, our findings suggest that while the framework is not yet fully practical in its current form, it provides (as expected) a credible and valuable foundation for the next stage in the evolution of scientific culture. In principle, it establishes a conceptual basis for greater transparency and accountability in sustainability reporting for research. However, widespread adoption will require the development of concrete tools, detailed guidance, and broadly accepted standards to translate this vision into everyday research practice.

\section{Limitations and future work}

The present validation is based on a relatively small expert sample and relies on self-reported perceptions rather than observational evidence of adoption. As such, the findings primarily reflect expert judgment on relevance and feasibility rather than demonstrated implementation outcomes. Future work should therefore focus on piloting FAIR+S in concrete research infrastructures, evaluating automated measurement and reporting tools, and refining minimal sustainability metadata requirements across domains. Longitudinal studies examining how sustainability reporting practices evolve under FAIR+S-aligned incentives, as well as alignment with journal, funder, and infrastructure policies, will be essential to support broader adoption.

In conclusion, the results indicate that FAIR+S addresses a recognised gap in current research data and software governance by embedding sustainability into the established FAIR paradigm. While further operationalisation is required, FAIR+S provides a validated and extensible foundation for advancing environmentally accountable, transparent, and reusable research practices.

\section{Concluding Remarks}

This study provides an empirical validation of FAIR+S as an extension of the FAIR and FAIR4RS principles that explicitly integrates sustainability considerations into research data and software practices. Based on a cross-disciplinary expert survey, the results demonstrate broad conceptual support for the FAIR+S principles, with experts consistently rating their importance as moderate to high across experience levels and domains.

The validation further indicates that FAIR+S is perceived as providing tangible added value rather than constituting an additional reporting burden. In particular, experts emphasised the importance of transparency, standardised sustainability benchmarks, and life-cycle sustainability disclosures as mechanisms to improve the trustworthiness, comparability, and interpretability of research artefacts. These findings suggest that sustainability-aware metadata and reporting are increasingly viewed as integral components of responsible and reproducible research.

At the same time, the study reveals a clear implementation gap. While the principles themselves are widely endorsed, their practical feasibility is constrained by the lack of standardised metrics, the limited availability of automated and workflow-integrated tools, and uncertainty regarding appropriate levels of reporting granularity. These challenges closely mirror earlier adoption barriers observed for FAIR, indicating that FAIR+S should be understood as an incremental extension that will require sustained community engagement and infrastructure support to mature.

From a broader conceptual perspective, FAIR+S could potentially be positioned within the framework of Environmental Management Accounting (EMA)\cite{Burritt2002}, which may offer a suitable foundation for integrating both physical sustainability indicators (e.g., kg CO$_2$ emissions) and monetary measures such as costs. However, exploring this connection lies beyond the scope of the present validation study and remains an avenue for future research.

\section*{Acknowledgements}
The authors thank the research communities and organisations that supported this work, particularly the Green Software Foundation (\url{https://greensoftware.foundation/}), the Software Sustainability Institute (\url{https://www.software.ac.uk/}), the NFDI4Energy community (\url{https://nfdi4energy.uol.de/}), and Wilhelm Büchner Hochschule (\url{https://www.wb-fernstudium.de/}) for their assistance in disseminating the survey and supporting participant recruitment. We are also grateful to the expert participants for sharing their insights and expertise. Special thanks to Dr. Fernando Pe\~naherrera, Dr. Ildar Baimuratov, Dr. Mirko Sch\"afer, Dr. Alexandra Pehlken, Viktor Dmitriyev, Tim Wittenborg, Philipp Weisenburger, Lea Kuhlmann, and Kirsty Pringle for their help and valuable feedback. Additionally, we thank the Energy-efficient Smart Cities group (Dr. Sven Rosinger, \url{https://www.offis.de/en/applications/energy/energy-efficient-smart-cities.html}) at OFFIS – Institute for Information Technology and the Very Large Business Applications group (Prof. Dr. habil. Jorge {Marx G\'omez}, \url{https://uol.de/vlba}) at Carl von Ossietzky Universit\"at Oldenburg for piloting the study and providing constructive feedback on early stage.

\section*{Author Contributions}
DV – Writing original draft, Review \& editing, Conceptualisation, Methodology, Data curation, Validation, Visualisation, Software; JSS – Writing original draft, Review \& editing, Data curation; JMG \& RI – Review \& editing, Conceptualisation, Supervision.

\section*{Competing interests}
The authors declare that they have no conflicts of interest regarding this research.

\section*{Data \& Code availability statement}
Data and Code are available in an open repository: \url{https://github.com/ellariel/fairs-pilot-validation}

\section*{Ethics Approval \& Informed Consent}
The study was approved by the Commission for Research Impact Assessment and Ethics at Carl von Ossietzky Universit\"at Oldenburg (Drs.EK/2025/103).

\section*{Funding Declaration}
This research received no particular funding.

\section*{Appendix A -- Expert Survey Questions}
\label{sec:appendixA}

\begin{longtable*}{|L{6.5cm}|L{8.5cm}|}
\hline
\textbf{Question} & \textbf{Scale / Answer Type} \\
\hline
\endfirsthead

\hline
\textbf{Question} & \textbf{Scale / Answer Type} \\
\hline
\endhead

\hline
\endfoot

\hline
\endlastfoot
 
\q{q01}Q\ref{q01}. What is your primary field of research or development?
& \vspace{0.09pt} Multiple choice: Computer Science related, Engineering related, Social Sciences related, Life Sciences related, Other
\\\hline
\q{q02}Q\ref{q02}. How many years of experience do you have in research?
& \vspace{0.09pt} Less than 5 / 5–10 / 11–15 / More than 15
\\\hline
\q{q03}Q\ref{q03}. What is your level of education or expertise?
& \vspace{0.09pt} Bachelor’s / Master’s / PhD / PhD+ (Prof., Group Lead, etc.)
\\\hline
\q{q04}Q\ref{q04}. How often do you participate in editorial boards, research or conference committees, or grant review activities that assess paper submissions or funding applications?
& \vspace{0.09pt} Never / Once every few years / Once a year / 2–3 times a year / More than 3 times a year
\\\hline
\q{q05}Q\ref{q05}. Do you have experience collecting data or preparing datasets for research purposes?
& \vspace{0.09pt} No experience / Limited experience (did it once) / Moderate experience / Extensive experience
\\\hline
\q{q06}Q\ref{q06}. Do you have experience developing software or using programming languages for research?
& \vspace{0.09pt} No experience / Limited experience (did it once) / Moderate experience / Extensive experience
\\\hline
\q{q07}Q\ref{q07}. How familiar are you with the FAIR principles (Findable, Accessible, Interoperable, Reusable)?
& \vspace{0.09pt} 1 (Not familiar) – 5 (Very familiar)
\\\hline
\q{q08}Q\ref{q08}. How important is it to follow FAIR principles?
& \vspace{0.09pt} 1 (Not important) – 5 (Extremely important)
\\\hline
\q{q09}Q\ref{q09}. To what extent do you apply the FAIR principles in your research or software development work?
& \vspace{0.09pt} 1 (Never apply) – 5 (Always apply)
\\\hline
\q{q10}Q\ref{q10}. How familiar are you with the sustainability concept (e.g., Sustainable Development Goals) in general?
& \vspace{0.09pt} 1 (Not familiar) – 5 (Very familiar)
\\\hline
\q{q11}Q\ref{q11}. How familiar are you with green software practices in general?
& \vspace{0.09pt} 1 (Not familiar) – 5 (Very familiar)
\\\hline
\q{q12}Q\ref{q12}. How important is it to include energy efficiency attributes in research artefact metadata?
& \vspace{0.09pt} 1 (Not important) – 5 (Extremely important)
\\\hline
\q{q13}Q\ref{q13}. Do you consider it practical to include energy efficiency attributes in research artefact metadata?
& \vspace{0.09pt} 1 (Strongly impractical) – 5 (Highly practical)
\\\hline
\q{q14}Q\ref{q14}. How feasible is it to report hardware/software/network context or specifications when publishing research?
& \vspace{0.09pt} 1 (Not feasible) – 5 (Very feasible)
\\\hline
\q{q15}Q\ref{q15}. Which hardware/software details do you currently report (if any)?
& \vspace{0.09pt} Multiple choice: CPU/GPU type, Memory, Software framework, Operating System / Environment, Cloud provider/region, System logs / Energy logs, None of them, Other
\\\hline
\q{q16}Q\ref{q16}. What challenges do you foresee in reporting energy efficiency attributes?
& \vspace{0.09pt} Open-ended
\\\hline
\q{q17}Q\ref{q17}. Should journals/conferences require some energy efficiency reporting?
& \vspace{0.09pt} Yes / No / Unsure
\\\hline
\q{q18}Q\ref{q18}. If yes, what form of reporting or which metrics would be most useful?
& \vspace{0.09pt} Open-ended
\\\hline
\q{q19}Q\ref{q19}. To what extent do you agree that validated sustainability benchmarks/tools are necessary?
& \vspace{0.09pt} 1 (Strongly disagree) – 5 (Strongly agree)
\\\hline
\q{q20}Q\ref{q20}. How likely would you be to adopt sustainability benchmarks if clear community standards existed?
& \vspace{0.09pt} 1 (Not likely) – 5 (Very likely)
\\\hline
\q{q21}Q\ref{q21}. Which methods would you most trust for energy consumption and carbon estimation?
& \vspace{0.09pt} Multiple choice: Direct measurement (e.g., power measurements, cloud billing data) / Community tools (e.g., CodeCarbon) / Simulation and Modelling estimates / Expert estimates
\\\hline
\q{q22}Q\ref{q22}. How valuable would standardise sustainability certifications (similar to FAIR badges) be for software and datasets?
& \vspace{0.09pt} 1 (Not valuable) – 5 (Extremely valuable)
\\\hline
\q{q23}Q\ref{q23}. What barriers might prevent you from adopting such frameworks and standards?
& \vspace{0.09pt} Open-ended
\\\hline
\q{q24}Q\ref{q24}. How important is transparency in sustainability reporting for research artefacts?
& \vspace{0.09pt} 1 (Not important) – 5 (Extremely important)
\\\hline
\q{q25}Q\ref{q25}. Would you support mandatory disclosure of methods/tools used for energy measurement?
& \vspace{0.09pt} Yes / No / Unsure
\\\hline

\q{q28}Q\ref{q28}. Would lifecycle sustainability statements improve your trust in published datasets/software?
& \vspace{0.09pt} Yes / No / Unsure
\\\hline

\q{q30}Q\ref{q30}. Which FAIR+S principles (S1–S5) do you find most critical for adoption?
& \vspace{0.09pt} Matrix scale: 1 (Not important) – 5 (Very important)

\\\hline
\end{longtable*}

\section*{Appendix B -- Expert Validation Methodology}
\label{sec:appendixB}

This study adopts a Design Science Research (DSR) approach to evaluate the proposed artefact (the FAIR+S framework) through a structured expert validation survey. In line with established DSR guidelines, artefacts must be rigorously evaluated in terms of their utility, quality, and applicability in a given context\cite{hevner2004design, Peffers2007}. Given the normative-driven and conceptual nature of FAIR+S, an ex-ante evaluation strategy based on expert judgment was selected as an appropriate method to assess its relevance, added value, and practical feasibility prior to large-scale implementation\cite{Venable2016}.

\subsection*{Expert Survey Design}

The survey instrument was developed following best practices in expert elicitation and survey methodology\cite{Rowe2001, Burgman2011, Bojke2021, Dillman2014}. The design aimed to ensure the survey content validity, clarity, and reliability, while minimising respondent burden. To achieve this, the questionnaire combined structured quantitative items with qualitative prompts, enabling both standardised measurement and in-depth insights\cite{Boone2012, Yan2012}.

The instrument was structured into three main components. First, a profiling section captured participants’ background, including research experience, education level, and domain expertise, ensuring that responses could be contextualised and filtered based on expertise. Second, baseline questions assessed familiarity with key concepts relevant to the artefact (e.g., FAIR principles, sustainability frameworks, and green software practices), establishing a foundation for interpreting subsequent evaluations. Third, the core evaluation section introduced the FAIR+S framework in a structured manner and elicited expert judgments across three dimensions that are also aligned with the theory of diffusion of innovations\cite{rogers2003diffusion} and generic technology acceptance model\cite{WARD2013}: \textit{perceived importance}, \textit{added value}, and \textit{feasibility}.

A variety of question formats were employed, including Likert-scale items, multiple-choice questions, matrix evaluations, and open-ended responses. Likert scales were used to capture perceptions and attitudes in a consistent and comparable manner, as recommended for attitudinal research\cite{Boone2012}, while open-ended questions supported the identification of barriers, enablers, and potential improvements, contributing to interpretive depth and triangulation\cite{Yan2012}.

Prior to deployment, the survey was piloted with a small group of experts to refine wording, improve clarity, and ensure appropriate length and comprehensibility. This iterative pretesting process is essential for reducing measurement error and enhancing reliability\cite{Dillman2014, Presser2004}.

\subsection*{Expert Sampling, Panel Size, and Recruitment}

Participants were recruited using a purposive sampling strategy, which is widely recommended for expert-based evaluations where domain knowledge is critical\cite{Rowe2001, Burgman2011}. The recruitment targeted researchers, data stewards, and software practitioners with relevant expertise in research data management, software development, sustainability, or governance.

A key methodological consideration in expert elicitation is the appropriate number of experts. Unlike statistical surveys aimed at population inference, expert studies prioritise quality and diversity of expertise over large sample sizes. The literature suggests that panels typically range between 10 and 30 experts, depending on the heterogeneity of the domain and the complexity of the subject matter\cite{Burgman2011, Bojke2021}. Smaller panels (e.g., 10–15 experts) may be sufficient when expertise is highly specialised and homogeneous, whereas more diverse or interdisciplinary topics benefit from somewhat larger panels to capture a broader range of perspectives\cite{Rowe2001}. Furthermore, studies on expert judgment aggregation indicate that reliability gains tend to diminish beyond moderate panel sizes, meaning that increasing the number of experts does not proportionally improve the quality of results\cite{Budescu2014}. In the context of design science evaluation, such panel sizes are considered adequate for assessing artefact relevance and feasibility, particularly when combined with rigorous selection criteria and mixed-method analysis\cite{Venable2016}. Therefore, the final expert sample in this study falls within recommended ranges and is deemed sufficient to support a robust and credible evaluation.

To ensure both expertise and diversity, invitations were disseminated through multiple channels, including academic networks, research institutions, professional organisations, mailing lists, conferences, and online platforms. This multi-channel approach helps mitigate selection bias and supports disciplinary diversity, which is important for robust expert elicitation\cite{Burgman2011}.

Following data collection, eligibility criteria were applied to identify a qualified expert panel. Participants were required to demonstrate either conceptual familiarity with relevant domains (e.g., FAIR principles or sustainability frameworks) or substantial professional experience (e.g., at least five years of research experience or a doctoral degree), in addition to practical experience in data or software-related work. This filtering step ensures that the final sample reflects epistemic authority and domain competence, which are critical for valid expert judgment\cite{Burgman2011, Budescu2014}. The resulting sample size is consistent with recommendations for expert elicitation studies, where smaller but well-qualified panels are preferred over larger, less specialised samples\cite{Bojke2021}.

\subsection*{Ethical Considerations}

The study adhered to established ethical standards for human subject research. Participation was voluntary, informed consent was obtained digitally, and responses were collected anonymously. Ethical approval was granted by the relevant institutional review body. These measures ensure compliance with research ethics principles, including confidentiality, autonomy, and data protection.

\subsection*{Data Collection Procedure}

The survey was administered over a defined period and distributed electronically. Clear instructions and contextual information were provided to ensure consistent understanding of the artefact being evaluated. The structured deployment, combined with targeted reminders where appropriate, follows recommended practices for maximising response quality and completion rates\cite{Dillman2014}.

\subsubsection*{Data Analysis}

A mixed-methods analytical approach was employed to provide a comprehensive evaluation of the artefact. Quantitative responses were analysed using descriptive statistics to identify central tendencies and variability, as well as non-parametric tests where appropriate to assess differences and agreement patterns. Such methods are widely used for analysing Likert-scale data and expert judgments\cite{Boone2012}.

In parallel, qualitative responses were analysed using content analysis to systematically identify recurring themes, including perceived barriers, enabling factors, and suggestions for improvement. This approach allows for structured interpretation of open-ended data and supports the identification of patterns not captured by quantitative measures\cite{Mayring2015}.

The integration of quantitative and qualitative findings enabled triangulation, enhancing the robustness and validity of the evaluation by combining statistical evidence with expert interpretation\cite{Yan2012, Steenbergen2007}.

\subsubsection*{Evaluation Logic within DSR}

The evaluation framework is explicitly aligned with DSR principles. The three core dimensions -- importance, added value, and feasibility correspond to key evaluation criteria in design science: relevance to the problem context, utility of the artefact, and practical applicability in real-world settings\cite{hevner2004design, Venable2016}. By systematically assessing these dimensions through expert judgment, the study provides evidence for the artefact’s validity and identifies areas requiring further refinement.

Thus, this methodology enables a rigorous, theory-informed, and empirically grounded validation of the proposed framework, supporting its iterative improvement and future implementation.

\printbibliography[heading=references]

\end{document}